\documentstyle[aps,prl,twocolumn]{revtex}
\input epsf
\begin{document}
\wideabs{
\title{Dynamic Fracture in Single Crystal Silicon} 
\author{Jens A.  Hauch, Dominic Holland, M. P. Marder, and Harry L. Swinney\\ 
Center for Nonlinear Dynamics and Department of 
Physics
\\ The University of Texas at Austin, Austin, TX 78712}
\date{\today}
\maketitle
\begin{abstract}
We have measured the velocity of a running crack in brittle single crystal
silicon as a function of energy flow to the crack tip. The experiments are designed
to permit direct comparison with molecular dynamics simulations; therefore the experiments provide
an indirect but sensitive test of interatomic potentials. Performing
molecular dynamics simulations of brittle crack motion at the atomic scale
we find that experiments and simulations disagree showing that interatomic
potentials are not yet well understood.
\end{abstract}
}


Data on fracture in single crystals are limited due to
the difficulties in performing precisely controlled experiments.
There is also a lack of atomic scale simulations that allow
quantitative comparison with fracture experiments.
We have obtained both experimentally and numerically 
the velocity $v$ of a crack propagating
in a silicon single crystal as a function of the energy
flux to the crack tip, the fracture energy $G$.
The relation between $v$ and $G$ is very sensitive to crystal structure
and details of interatomic forces. Thus the
experimentally determined $v(G)$ provides a test
of the interatomic potentials used in simulations.
We find poor quantitative agreement between simulation 
and experiment, showing that the
existing potentials do not capture the complexities of fracture.

\textit{Experiments} --
We chose silicon for our experiments 
since it is very brittle at room temperature \cite{lawn} 
and readily available as oriented single crystals.
The lowest energy cleavage plane in silicon is the \{111\} plane. All experiments reported
here were performed on p-type \{110\} wafers (0.38~mm thick) with a doping level of $\sim 10^{19}$ boron atoms
per $\rm cm^3$; \{110\} wafers are the only commercially available wafers with a \{111\}
plane normal to the plane of the wafer.

\begin{figure}[!ht]
\smallskip
\centerline{\epsfxsize=3.1truein\epsffile{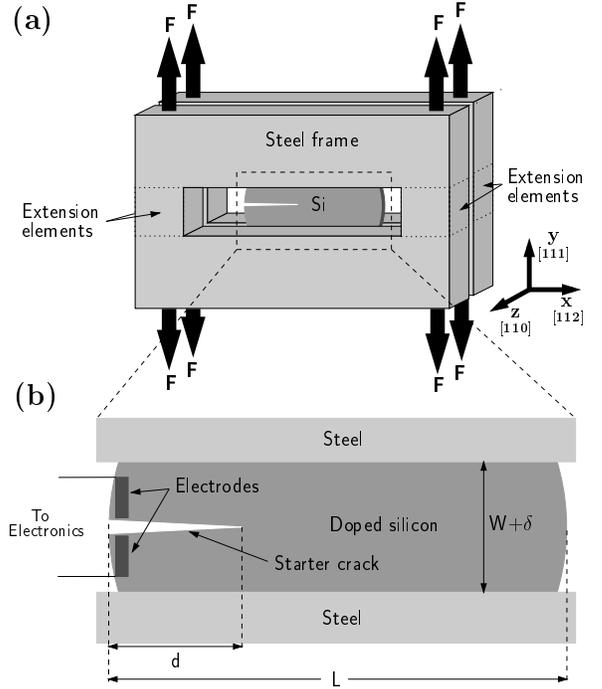}}
\caption{(a) Rigid jaw loading configuration. The steel frames 
are loaded symmetrically with a constant force at 8 points.
This causes a small extension of the extension elements, 
leading to a small displacement $\delta$
of the inside edges of the frame. (b)  A wafer 
in the frame. The extension of the frame enforces
constant displacement boundary conditions on the wafer. }
\label{loading}
\end{figure}

Previous experiments in single crystal silicon \cite{ericson3,gilman2,chen} 
measured the minimum
energy density required to drive a crack,
and a few have measured dynamic crack behavior
\cite{brede1,cramer}, but without the ability
to measure the fracture energy. In our experiments
samples were
loaded in a thin strip configuration by displacing the edges
of the wafer a constant amount $\delta$, as shown in 
Fig.~\ref{loading} (b). Thin strip refers to samples
with an aspect ratio $L/W > 1$. In all experiments described 
here, $L=7.5$ cm and $W=3.3$ cm, giving an aspect ratio of 2.3.
The advantage of this tensile loading geometry is that 
if the boundaries of the sample can be held fixed
while the crack propagates, then
the energy released to the crack 
tip is independent of crack length and
results in steady state fracture at constant 
velocity \cite{rice,holland,gumbsch2}.
In this loading configuration, $G$ is simply the
elastic strain energy stored per unit area (xz-plane in Fig.~\ref{loading}~(a))
ahead of the crack, and can be written as a simple function of the
sample width $W$, width extension $\delta$,
Young's modulus $E$, and for very thin samples,
giving plane stress conditions, Poisson's ratio $\nu$:
\begin{equation}
\label{G}
G = {1\over 2}{E\over{1-\nu^2}}{{\delta^2}\over W}.
\end{equation}

To open a crack along a \{111\} plane, 
$\delta$ must be about 1~$\mu$m ($E_{111}=188 \rm ~GPa$, $\nu$=0.272 \cite{brantley},
$G=2.4~ \rm J/m^2$ \cite{spence}).
In a 
controlled experiment $\delta$ 
has to be maintained constant
along the whole length of the wafer.
The control required
to maintain small, constant displacements
is difficult to achieve in a conventional tensile testing machine
because as a sample fractures, the force it
exerts on the testing machine decreases
and the machine responds rapidly enough to
affect the crack dynamics.
This effect is independent of
the stiffness of the testing machine \cite{hauch}. 
Therefore we designed the frame loading configuration shown in
Fig.~\ref{loading}~(a). It consists of two steel 
frames with a rectangular hole milled out of each. In an experiment,
the silicon sample is clamped and glued with slow curing 
cyanoacrylate adhesive between the two frames, exposing a thin
strip in the hole. When loaded, the frames act
as two rigid bars connected by two extension elements which function
as very stiff springs. 

To extend the extension elements by 1 $\mu$m
requires a loading force of about $\rm8000\, N$ 
distributed over the eight loading points. 
The extension pulls apart  
the inside edges of the hole, which in turn enforces a
constant displacement on the edges of
the silicon wafer. Fracture of the sample does not 
lead to a relaxation of the frame, 
since only a small
fraction ($< 2$\%) of the total load is transmitted through the silicon sample.
One concern, however, is non-ideal deformation
of the rigid frame itself. Finite element analysis of 
the sample and frame
shows that the total displacement of the
edges of the wafer is constant to within 10\% along 
the length of the sample.

\begin{figure}[tbp]
\smallskip
\centerline{\epsfxsize=3.3truein\epsffile{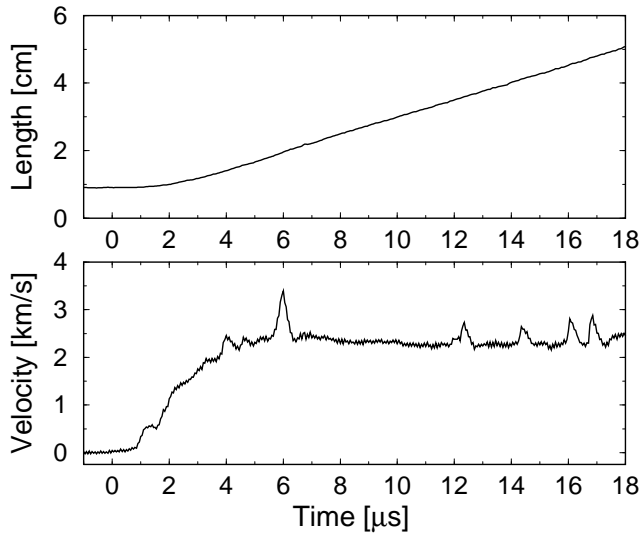}}
\caption{Full length and velocity record for a break in 
silicon along the \{111\} plane in the $[\bar{1} \bar{1} 2]$ 
direction at a fracture energy of $\rm5.08\, J/m^2$.  
The velocity peaks that
are visible are probably the result of acoustic waves, 
generated by the crack initiation, interacting with the
crack tip.}
\label{vellen}
\end{figure}

To obtain fracture along a \{111\} plane, it is necessary 
to start with a seed crack in this plane.
The seed crack is formed by a
thermal shock technique involving three steps.
First, the sample is notched with a diamond disc saw. Next,
the sample is heated by dipping it partially into boiling water.
Finally, the hot sample is rapidly lowered to the desired
seed crack length (1 to 3 cm) into
ice-water so that the thermal stresses
induced in the sample provide an opening force 
perpendicular to the \{111\} plane, resulting in a
sharp seed crack. Length and sharpness of the seed crack determine
the failure stress of the sample. Therefore by varying the seed crack 
length it is possible to get cracks to propagate at different fracture energies.

To measure crack length and velocity during crack propagation,
a potential-drop technique similar to the one outlined 
in ref.\cite{hauch} was used.
Potential-drop techniques monitor the change in resistance of a conductor,
in this case doped silicon, during crack propagation. Resistance
was measured by attaching electrodes on both sides
of the seed crack as shown in Fig.~\ref{loading} (b). The 
electrodes were then connected to a Wheatstone bridge. 
To interpret the data, a lookup table of resistance as a 
function of crack length was created.
The bridge output was digitized at $\rm10\,MHz$ (12 bit), leading 
to a resolution
of about $\rm 1\, mm$ in crack length. To attain good
resolution of the velocity, the output of the bridge was fed into an analog 
differentiator and digitized at $\rm20\, MHz$ (8 bit). With this method the velocity 
measurements were limited by noise to a
resolution of $\rm 50\, m/s$. A full data set obtained by this method is shown 
in  Fig.~\ref{vellen}.
The data show that after an initial acceleration stage, the crack 
velocity settles into a steady value of $\rm 2.3 \pm 0.3\, km/s$. The 
experimentally observed crack velocities range from $\simeq$40-75\% of
the transverse sound speeds in silicon in agreement with previously observed
crack velocities \cite{field}.

\begin{figure}[htbp]
\smallskip
\centerline{\epsfxsize=2.8truein\epsffile{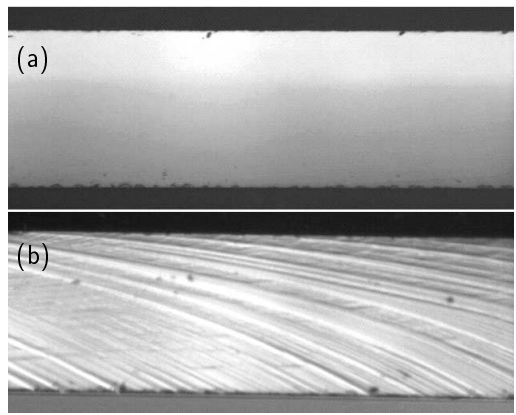}}
\vspace{0.1truein}
\caption{
(a) Featureless fracture  surface resulting from uniform loading; (b) surface structure resulting
from bending of the sample during the experiment. The region shown is $\rm1.6\,mm \times 0.38\,mm$.
}
\label{surf}
\end{figure}

\begin{figure}[htbp]
\smallskip
\centerline{\epsfxsize=3.3truein\epsffile{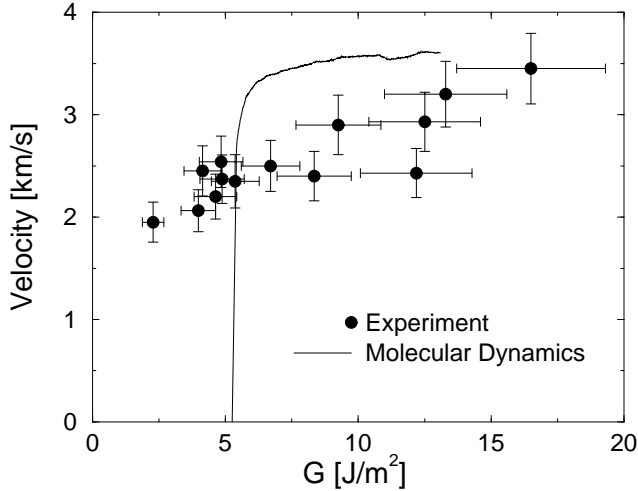}}
\caption{Crack velocity as a function of fracture 
energy from experiments and molecular dynamics simulations at 300 K. 
The error bars in the experiments are the result of
random errors due to variations in sample resistivity, 
while the uncertainty in the fracture energy is systematic and the
result of uncertainty in the strain obtained from
strain gage measurements. All low-lying velocity states
exist in the simulations for temperatures down to about 200~K.
}
\label{vvsg}
\end{figure}

Samples were loaded
quasi-statically by incrementing $\delta$ by about $0.03~\mu$m
at intervals of $60$~s. 
The majority of experimental attempts had to be discarded 
because nonuniform curing of the adhesive caused the wafers
to bend slightly out of plane during loading,
so that the samples were twisted, not just stretched.
Such wafers were easily identified by
examining the surface created by the passing crack.
If a bending moment existed during fracture, the crack
would leave behind structure on the fracture surface, as 
shown in Fig.~\ref{surf}~(b). Under these conditions Eq.~(1) would not apply and
all the samples that displayed these markings were 
eliminated, while only samples that showed a perfect
mirror-smooth fracture surface, as shown in 
Fig.~\ref{surf}~(a), were kept. 
The plot of average crack velocity as a function 
of fracture energy for all the remaining samples
is shown in Fig.~\ref{vvsg}.

\textit{Simulations} --
We have also carried out molecular dynamics simulations of fracture in silicon,
the results of which are also shown in Fig.~\ref{vvsg}.
The simulations  are carefully designed to measure
numerically the same quantities measured in experiments,
despite the great difference in scale between them.
Since the experiments are performed at room temperature,
the simulations were also maintained at 300 K by contact with
a heat bath.
These simulations, which are described in Ref.~\cite{holland,hollandaise}, 
use a modified Stillinger-Weber interatomic potential. The original
Stillinger-Weber potential \cite{stillinger}, like the more 
sophisticated environment-dependent 
interatomic potential \cite{Bazant.97.prb,bazant3}, did not yield 
brittle crack propagation. 
Most potentials for silicon, of which there are over 30 \cite{balamane},
have a range restricted to nearest neighbors, $\simeq 3.5$~\AA,
although density functional theory predicts a range of $\simeq 5.5$~\AA.
Because of the short cutoff, potentials must rise from the
cohesive well and go to zero rapidly,
resulting in an unreasonably large force of attraction
before rupture. This large force inhibits crack propagation:
the seed crack tip blunts and will not move; at very high strains,
the tip simply melts.
Without changing the form of the Stillinger-Weber potential, we were able to
get cracks propagating along the experimentally preferred
fracture planes, \{111\} and \{110\}, simply by
doubling the strength of the term enforcing fixed angles
between bonds. With this modified potential, it is possible
to observe brittle fracture and calculate the relation between 
crack velocity $v$ and fracture energy $G$. 
While this modification does not affect the Griffith point and 
makes a quantitative comparison to the experiments possible
the changed potential cannot be regarded as superior to the original Stillinger-Weber
potential since some of the material properties are changed.
The modification changes the melting temperature from $\simeq 1400$~K to $\simeq 3500$~K
compared to $\simeq 1685$~K for real silicon. The elastic properties are also
affected by the modification; for example Young's modulus along the $[111]$
direction is changed from $E_{111}=151$ GPa to $E_{111}=207$ GPa for the modified potential, 
while experimentally $E_{111}=188$ GPa.
Improved interatomic potentials are clearly needed.

The results from simulations presented in Fig.~\ref{vvsg}
were obtained for a
thin strip of size
$532\times 15\times 154~ \mbox{\AA}^3$, periodic along
the thin axis. New material was
added ahead of the crack tip and old material lopped off at the tail
every time the crack advanced to within 200~\AA~of the forward end of the strip.
In this fashion, the crack traveled 7 $\mu$m during the course of
the simulations as $G$ was varied between
5 and 14 J/$\mbox{m}^2$.

\textit{Comparison} -- 
The highest crack velocities observed in Fig.~\ref{vvsg} are
reasonably close, but the minimum fracture energies
at which a crack propagates differ: 2.3~$\rm J/m^2$ in the experiments and
5.2~$\rm J/m^2$ in the simulations.
Since the scale of crack velocities in a material is bounded
by sound speeds \cite{freund-book}, which potentials get approximately right,
it is not surprising that the experimental and computational crack velocity
scales agree. Furthermore, the potential gives the correct
cohesive energy of silicon, leading to the agreement in numerical 
and experimental energy scales.
However, the nonlinear
parts of the potential involved in stretching and
rupturing bonds play an important role in determining the actual
fracture energies and crack velocities, in particular
where the crack arrests and what its highest velocity is.
The quantitative disagreements we find point to a
shortcoming of the nonlinear parts of the potential,
which have not received much attention.

The lowest fracture energy at which a crack propagated in the experiments 
was close to the lower bound estimate of the
fracture energy for a \{111\} plane, 2.2~J/$\mbox{m}^2$ \cite{spence},
which is twice the surface
energy density. Since a crack cannot
travel with less energy, there must be a narrow range
of fracture energy over which the crack velocity rises rapidly from 
zero to the lowest value measured, $\simeq 2$~km/s.
This phenomenon is also seen in glass and
polymers and is reminiscent of
a velocity gap in the lattice models of Marder et al. \cite{marder3}
and in simulated systems \cite{holland,gumbsch2}.
A velocity gap is a band of velocities in which
it is impossible for a crack to travel in steady state.
In our numerical silicon, the velocity gap is temperature dependent,
vanishing above 200 K.
At 300 K there do exist steady states at all velocities
between 0 and 3 km/s. 

Steady state experiments are difficult in the narrow 
range of fracture energy where
$v(G)$ rises sharply 
because of the extreme precision required at the
boundaries. However, in glass and Plexiglas it was possible 
to rule out the existence 
of a velocity gap by conducting carefully controlled
crack arrest experiments, which showed,  by slowly decreasing the
energy flux to the crack tip, that all velocities are 
available to the crack \cite{hauch}.
A velocity gap 
should show up as a very rapid velocity jump between 
zero and roughly 20-40\% of the sound speed 
at initiation.
The accelerations observed in the experiments can be
compared to the timescales available to the system through the 
nondimensional acceleration  $\bar{a}= a W/c^2$, 
where $c$ is the sound velocity, $W$ the sample width, and $a$
is the acceleration. 
In our experiments we routinely observe accelerations on the 
order of $\rm10^9~m/s^2$ in silicon, which corresponds
to $\bar{a} \simeq 1$. This acceleration is not large, 
but due to three dimensional averaging
over the whole crack front that is involved in our measurement,
our data do not completely rule out
the possibility of a velocity gap in silicon at room temperature.

The experiments covered a range of fracture energies, $\rm2-16~ J/m^2$,
in which the cracks produced very smooth surfaces. 
Thus cracks can dissipate large amounts of energy, more
than seven times the amount needed to create a clean 
cleavage through the whole crystal, 
without leaving behind any large scale damage on 
the fracture surfaces. Investigation by atomic force
microscopy shows that for low fracture energies the 
fracture surface is atomically flat, while at higher
energies the surface has definite features. These 
features are smooth on the submicron scale, and account for
height variations on the order of $\rm30\, nm$ over 
an area of $\rm16\, \mu m^2$. 
The roughness gives an area increase of only $\sim$0.1\%
above that of a flat cleaved surface. This added surface cannot
account for the sevenfold increase in dissipated energy.
We do not know the mechanism by which the extra energy is dissipated.
However, the computations indicate that
most of it is carried off in lattice vibrations.
In polymers energy is also dissipated through the creation of small
micro-cracks, which are visible on the fracture surface in the form
of a mirror-mist-hackle transition \cite{sharon1,sharon2,ravi-chandar3,hauch}. 
The observation of such a transition
in silicon \cite{cramer} implies that 
a similar mechanism may be important in silicon as well.

\textit{Conclusions} --
We have measured the dependence of crack velocity on fracture energy in 
single crystal silicon.
We control the velocity of a running crack by controlling the energy flux to the 
crack tip. 
Experiments and simulations agree qualitatively, for both show an initial sharp rise in velocity,
followed by slowly increasing crack velocities as fracture energy increases. 
However, details in the relation between $v$ and $G$ are quite sensitive to details of
interatomic potentials. 
The ability to compare experiment and computation provides a strong test of 
potentials, but the quantitative disagreement we find demonstrates that they
are not yet correct.

This work was supported by the National Science 
Foundation (DMR-9802562, DMR-9531187),
the Texas Advanced Research Program,
the Texas Advanced Computing Center,
the National Partnership for Advanced Computational Infrastructure,
and the Exxon Education Foundation. We thank W. D. McCormick for advice
on innumerable technical issues, J. Hanssen and R. D. Deegan for a datum, 
C. K. Shih and R. Mahaffy
for performing atomic force microscopy, and G. Rodin for access
to finite element software.

\vspace{-1mm}


\vfill
\eject

\end{document}